\documentclass[preprint]{iucr}

\usepackage[latin1]{inputenc}
\usepackage{amssymb}
\usepackage{amsmath}
\usepackage{amsthm}
\usepackage{bmpsize}
\usepackage{color}
\usepackage[english]{babel}
\usepackage{exscale}
\usepackage{graphics}
\usepackage{graphicx}
\usepackage{harvard}
\usepackage{longtable}
\usepackage{mathtools}
\usepackage{pdflscape}
\usepackage{relsize}
\usepackage{soul}
\usepackage{tabularx}

\begin{document}

\title{Second order superstructure reflexions in tilted perovskites}
\shorttitle{Octahedral tilting rules}

\author[a]{Richard}{Beanland}
\cauthor[a]{Robin}{Sj\"okvist}{Robin.Sjokvist@warwick.ac.uk}

\aff[a]{Department of Physics, University of Warwick, \city{Coventry} CV4 7AL, \country{UK}}

\maketitle

\begin{abstract}
In a previous article \cite{B+S}, we derived Boolean conditions for the appearance of superstructure reflexions in diffraction patterns from perovskites with tilted oxygen octahedra, using the structure factor equation.  Assuming that the deviation from the untilted prototype perovskite structure was infinitesimally small, we expanded the structure factor as a Taylor series, truncated at the first term.  This gave an elegant and simple method giving conditions on the presence or absence of superstructure reflexions.  However, in real perovskite materials these distortions are sufficiently large for higher order terms to become significant, giving an additional set of superstructure reflexions.  Here, we consider the second term in the expansion and show that it gives rise to `second order' reflexions with indices of the form $\frac{1}{2}$ even-even-odd.  Boolean conditions for their presence are given.
\end{abstract}

\newpage

\section{Introduction}

~

In \cite{B+S}, we derived Boolean conditions for the appearance of superstructure reflexions in tilted perovskites, taking the displacements of the oxygen atoms from their positions in the prototype unit cell to be $\boldsymbol{\delta}^{(j)}$.  In this framework, the structure factor $F_g$ for a reflexion $\textbf{g}$ with Miller indices $g_1~g_2~g_3$ (or $hkl$ in the convention used in the International Tables for Crystallography,\cite{IntTables}) can be written

\begin{equation}
    \label{Struc}
    F_g = \sum_{j=1}^{n} f_g^{(j)} \exp{\left(2 \pi i \textbf{g} \cdot \textbf{r}^{(j)} \right)}\exp{\left(2 \pi i \textbf{g} \cdot \boldsymbol{\delta}^{(j)}\right)}
\end{equation}

where the sum is taken over all $j$ atoms in the unit cell, each having atomic scattering factor $f_g^{(j)}$ and fractional coordinates $\textbf{r}^{(j)}$ in the prototype structure.  A simple equation giving the type of reflexions produced by a given tilt system was derived by expanding the last term as a series 

\begin{equation}
    \label{expan}
    \exp{\left(2 \pi i \textbf{g} \cdot \boldsymbol{\delta}^{(j)}\right)} = 1 + 2 \pi i \textbf{g} \cdot \boldsymbol{\delta}^{(j)} - 4 \pi^2 (\textbf{g} \cdot \boldsymbol{\delta}^{(j)})^2 + ...
\end{equation}

which can be truncated after the first-order term, if the displacements $\boldsymbol{\delta}^{(j)}$ are taken to be infinitesimally small, which was the premise of our previous article.  However, experimental electron diffraction patterns show additional sets of weak superstructure spots that are not given by the previous analysis, indicating that oxygen displacements in tilted perovskites are sufficiently large for second-order terms to be important.  These additional spots were termed `concert' reflexions by \cite{Woodward}.  Here, we derive equations for these sets of weak second-order superstructure reflexions.  We find that their type is complementary to first-order superstructure reflexions, and that they are only present in perovskites with oxygen octahedral tilts about more than one axis, at least one of which must be an in-phase tilt system.

\section{Calculation}

~

From the last term in Eq.~\ref{expan} and following the approach used in \cite{B+S} the structure factor of a second-order superstructure reflexion it produces $\textbf{g}$ is

\begin{equation}
    \label{SuperStruc}
    F_g = -\sum_{j=1}^{n} f_g^{(j)} 4 \pi^2 (\textbf{g} \cdot \boldsymbol{\delta}^{(j)})^2 A^{s^{(j)}_1}B^{s^{(j)}_2}C^{s^{(j)}_3},
\end{equation}

where $A$, $B$ and $C$ take values of $\pm 1$ or $\pm i$ for even or odd $g_i$ respectively, and the position of oxygen atoms is written as $\textbf{s}^{(j)} = 4\textbf{r}^{(j)}$.  Importantly, when Eq.\ref{SuperStruc} is evaluated for a specific tilt system, terms appear that are only non-zero for even, or odd, reflexion indices; for example, $(1 + A^2)$ is only non-zero when $g_1$ is even, while $(1 - B^2)$ is only non-zero when $g_2$ is odd.  A unit cell doubled in all three dimensions is used, large enough to be a unit cell for any Glazer octahedral tilting pattern, which contains 24 oxygen atoms and hence gives 24 terms in Eq.~\ref{SuperStruc}.  The doubling of the unit cell in the calculation means that, in the reference frame of the prototype perovskite cell, superstructure reflexions have half-order indices and, for example, an equation of the form $(1 - A^2)(1 - B^2)(1 + C^2)$ indicates that superstructure spots must have the form $hkl = \frac{1}{2} odd-odd-even$, which is shortened to $\frac{1}{2}ooe$.

It is helpful to split Eq.\ref{SuperStruc} into two by expanding the $\textbf{g} \cdot \boldsymbol{\delta}^{(j)}$ term, i.e. 

\begin{equation}
    \label{squar}
    (\textbf{g} \cdot \boldsymbol{\delta}^{(j)})^2 = (g^2_1\delta^2_1 + g^2_2\delta^2_2 + g^2_3\delta^2_3) + 2(g_1g_2\delta_1\delta_2 + g_2g_3\delta_2\delta_3 + g_3g_1\delta_3\delta_1),
\end{equation}

and we will consider the two terms in brackets in this equation separately.  First, we have 

\begin{equation}
    \label{SuperNull}
    F_g = -\sum_{j=1}^{n} f_g^{(j)} 4 \pi^2~(g^2_1\delta^2_1 + g^2_2\delta^2_2 + g^2_3\delta^2_3)~A^{s^{(j)}_1}B^{s^{(j)}_2}C^{s^{(j)}_3}.
\end{equation}

which is, substituting the 24 oxygen coordinates $\textbf{s}^{(j)}$

\begin{equation}
    \label{nullsum1}
    \begin{split}
    \frac{F_g}{2 \pi^2 f_g} 
    =~& -(g^2_1\delta^2_1 + g^2_2\delta^2_2)\\ &~~~(AB+AB^3+A^3B+A^3B^3+ABC^2+AB^3C^2+A^3BC^2+A^3B^3C^2)\\
    -~& (g^2_2\delta^2_2 + g^2_3\delta^2_3)\\ &~~~(BC+BC^3+B^3C+B^3C^3+A^2BC+A^2BC^3+A^2B^3C+A^2B^3C^3) \\
    -~& (g^2_1\delta^2_1 + g^2_3\delta^2_3)\\ &~~~(AC+AC^3+A^3C+A^3C^3+AB^2C+AB^2C^3+A^3B^2C+A^3B^2C^3)
    \end{split}
\end{equation}

and this reduces to

\begin{equation}
    \label{nullsum1a}
    \begin{split}
    \frac{F_g}{2 \pi^2 f_g} = &
    -\left((g^2_1\delta^2_1 + g^2_2\delta^2_2)AB+(g^2_2\delta^2_2 + g^2_3\delta^2_3)BC +(g^2_1\delta^2_1 + g^2_3\delta^2_3)AC \right)\\ &~~~(1+A^2)(1+B^2)(1+C^2)
    \end{split}
\end{equation}

which is only non-zero for $hkl = \frac{1}{2}eee$, i.e. $hkl$ are all integers and are thus coincident with the matrix reflexions.  In other words, no new set of superstructure reflexions is produced by this term.  The second part of Eq.\ref{squar} is

\begin{equation}
    \label{SuperSec}
    F_g = -\sum_{j=1}^{n} f_g^{(j)} 4 \pi^2~(g_1g_2\delta_1\delta_2 + g_2g_3\delta_2\delta_3 + g_3g_1\delta_3\delta_1)~A^{s^{(j)}_1}B^{s^{(j)}_2}C^{s^{(j)}_3}.
\end{equation}

This equation indicates that weak second-order superstructure reflexions are only produced by perovskites with at least two tilt systems, since all terms contain products of two different tilt systems $\delta_i\delta_j$, $i\ne j$.  Furthermore, it is found that any system that is comprised only of anti-phase tilts just produces terms containing $(1+A^2)(1+B^2)(1+C^2)$ which, like the example considered above, are coincident with matrix reflexions.  To summarise this point: no second-order superstructure reflexions are produced in the seven Glazer tilt systems $a^0a^0c^+$, $a^0a^0c^-$, $a^0b^-b^-$, $a^0b^-c^-$, $a^-a^-a^-$, $a^-b^-b^-$ and $a^-b^-c^-$.  They are found in the seven tilt systems $a^+a^+a^+$, $a^0b^+b^+$, $a^+b^+c^+$, $a^0b^+c^-$, $a^+a^+c^-$, $a^+b^-b^-$ and $a^+b^-c^-$.  (Note that, following \cite{H+S}, we do not consider tilt systems with a space group that coincides with one of the fourteen listed above, since they produce no additional reflexions.)  The results are summarised below in Table \ref{summary}, including the first-order superstructure reflexions derived in \cite{B+S}.

It is also instructive to examine particular examples.  One of the tilt systems that satisfies the minimum requirements for the presence of second-order superstructure reflexions is $a^0b^+c^-$, which has $\frac{1}{2}ooo$ first-order superstructure reflexions that result from the antiphase $c^-$ tilts and $\frac{1}{2}oeo$ reflexions from the in-phase $b^+$ tilts \cite{B+S}.  The second-order superstructure reflexions are of the type $\frac{1}{2}eoe$, i.e. complementary to the first-order $\frac{1}{2}oeo$ reflexions.  Similarly, the tilt system $a^+a^+c^-$ has first-order superstructure reflexions of the type $\frac{1}{2}eoo$, $\frac{1}{2}oeo$ and $\frac{1}{2}ooo$ and second-order superstructure reflexions of the type $\frac{1}{2}oee$ and $\frac{1}{2}eoe$.  That is, in general, any odd-odd-even first-order reflexion produced by in-phase tilts produces a complementary even-even-odd reflexion \emph{in the presence of an antiphase tilt}.

There are three structures with multiple in-phase tilt systems and no anti-phase tilts, i.e. $a^0b^+b^+$, $a^+a^+a^+$ and $a^+b^+c^+$.  In these structures, no even-even-odd second-order reflexions are produced and for the latter two cases the second order reflexions simply coincide with first-order $\frac{1}{2}ooe$, $\frac{1}{2}oeo$ and $\frac{1}{2}eoo$ reflections, but with fewer systematic absences.  For the $a^0b^+b^+$ tilt system, second-order superstructure reflections take the form $\frac{1}{2}eoo$ while first-order superstructure reflections have indices $\frac{1}{2}oeo$ and $\frac{1}{2}ooe$.  In this case we would therefore expect all three types of odd-odd-even reflexion to be present, but with the second-order reflexions having a significantly weaker intensity.\\

\begin{table}
    \caption{Pseudo-cubic second-order superstructure reflections in perovskites with octahedral tilting. Miller indices are given in the form $\textbf{g}=hkl$.}
    \label{summary}
        \begin{tabular}{m{0.05\linewidth} | m{0.1\linewidth} | m{0.1\linewidth} | m{0.45\linewidth} | m{0.25\linewidth}}
            No. & Tilt \mbox{system} & Space group & First order & Second order\\
            \hline
            1 &\(a^+a^+a^+\) & \(Im\bar{3}\) &
           \begin{tabular}[c]{@{}l@{}}\(\frac{1}{2}ooe,~|h| \neq |k|;\)\\\(\frac{1}{2}oeo,~|h| \neq |l|;\)\\\(\frac{1}{2}eoo,~|k| \neq |l| \)\end{tabular} &
            \begin{tabular}[c]{@{}l@{}}\(\frac{1}{2}ooe,~h,k \neq 0;\)\\\(\frac{1}{2}oeo,~h,l \neq 0;\)\\\(\frac{1}{2}eoo,~k,l \neq 0 \)\end{tabular} \\
            ~ & ~ & ~ & ~\\%\hline
            2 & \(a^0b^+b^+\) & \(I\frac{4}{m}mm\) & \begin{tabular}[c]{@{}l@{}}\(\frac{1}{2}ooe,~|h| \neq |k|;\) \\ \(\frac{1}{2}oeo,~|h| \neq |l|\)\end{tabular} & \(\frac{1}{2}eoo,~k,l \neq 0\) \\
            ~ & ~ & ~ & ~\\%\hline
            3 &\(a^0a^0c^+\) & \(P\frac{4}{m}bm\) & \(\frac{1}{2}ooe,~|h| \neq |k| \) & None \\
            ~ & ~ & ~ & ~\\%\hline
            4 & \(a^0a^0c^-\) & \(I\frac{4}{m}cm\) & \(\frac{1}{2}ooo,~|h| \neq |k| \) & None \\
            ~ & ~ & ~ & ~\\%\hline
            5 & \(a^0b^-b^-\) & \(Imma\) & \begin{tabular}[c]{@{}l@{}}\(\frac{1}{2}ooo,~!(|h|=|k|=|l|),~k \ne l, \)\\!\((h=n,~k=\pm n+4m,~l=\mp n+4m),\)\\ \(n, m\) integers\end{tabular} & None \\
            ~ & ~ & ~ & ~\\%\hline
            6 & \(a^-a^-a^-\) & \(R\bar{3}c\) & \begin{tabular}[c]{@{}l@{}}\(\frac{1}{2}ooo,~h \ne k~\&~k \ne l~\&~l \ne h, \)\\!\((h=n,~k=n+4m,~l=n+4p),\)\\ \(n, m, p\) integers\end{tabular} & None \\
            ~ & ~ & ~ & ~\\%\hline
            7 & \(a^+b^+c^+\) & \(Immm\) & \begin{tabular}[c]{@{}l@{}}\(\frac{1}{2}ooe,~|h| \neq |k|;\)\\\(\frac{1}{2}oeo,~|h| \neq |l|;\)\\\(\frac{1}{2}eoo,~|k| \neq |l| \)\end{tabular} & \begin{tabular}[c]{@{}l@{}}\(\frac{1}{2}ooe,~h,k \neq 0;\)\\\(\frac{1}{2}oeo,~h,l \neq 0;\)\\\(\frac{1}{2}eoo,~k,l \neq 0 \)\end{tabular} \\
            ~ & ~ & ~ & ~\\%\hline
            8 & \(a^+a^+c^-\) & \(P\frac{4_2}{n}mc\) & \begin{tabular}[c]{@{}l@{}}\(\frac{1}{2}ooo,~|h| \neq ~|k|;\)\\\(\frac{1}{2}oeo,~|h| \neq ~|l|;\)\\\(\frac{1}{2}eoo,~|k| \neq ~|l| \)\end{tabular} & \begin{tabular}[c]{@{}l@{}}\(\frac{1}{2}ooe,~h,k \neq 0;\)\\\(\frac{1}{2}oeo,~k,l \neq 0;\)\\\(\frac{1}{2}oee,~h,l \neq 0 \)\end{tabular} \\
            ~ & ~ & ~ & ~\\%\hline
            9 & \(a^0b^+c^-\) & \(Cmcm\) & \begin{tabular}[c]{@{}l@{}}\(\frac{1}{2}ooo,~|h| \neq |k|;\)\\\(\frac{1}{2}oeo,~|h| \neq |l| \)\end{tabular} & \(\frac{1}{2}eoe,~k,l \neq 0\) \\
            ~ & ~ & ~ & ~\\%\hline
            10 & \(a^+b^-b^-\) & \(Pnma\) & \begin{tabular}[c]{@{}l@{}}\(\frac{1}{2}eoo,~|k| \neq |l|;\)\\\(\frac{1}{2}ooo,~!(|h|=|k|=|l|),~k \ne l, \)\\!\((h=n, k=\pm n+4m, l=\mp n+4m),\)\\ \(n, m\) integers\end{tabular} & \(\frac{1}{2}oee,~h,k,l \neq 0\) \\
            ~ & ~ & ~ & ~\\%\hline
            11 & \(a^0b^-c^-\) & \(C\frac{2}{m}\) & \(\frac{1}{2}ooo, ~!(|h|=|k|=|l|)\) & None \\
            ~ & ~ & ~ & ~\\%\hline
            12 & \(a^-b^-b^-\) & \(C\frac{2}{c}\) & \(\frac{1}{2}ooo, ~!(|h|=|k|=|l|), ~k \neq l\) & None \\
            ~ & ~ & ~ & ~\\%\hline
            13 & \(a^+b^-c^-\) & \(P\frac{2_1}{m}\) & \begin{tabular}[c]{@{}l@{}}\(\frac{1}{2}ooo, ~!(|h|=|k|=|l|);\)\\\(\frac{1}{2}eoo,~|k| \neq |l| \)\end{tabular} & \(\frac{1}{2}oee,~h,k,l \neq 0\) \\
            ~ & ~ & ~ & ~\\%\hline
            14 & \(a^-b^-c^-\) & \(P\bar{1}\) & \(\frac{1}{2}ooo, ~!(|h|=|k|=|l|)\) & None \\
        \end{tabular}
\end{table}

\section{Results}

\begin{figure}
    \caption{Selected area electron diffraction patterns (SADPs) of a) $\langle1\bar{1}0\rangle$ and b) $\langle111\rangle$ Ba$_{0.78}$Ca$_{0.22}$TiO$_3$.  Data collected using a JEOL 2100plus high contrast TEM, operating at 200 kV.}
    \label{SCT1}
    \centering \includegraphics[width=0.8\columnwidth]{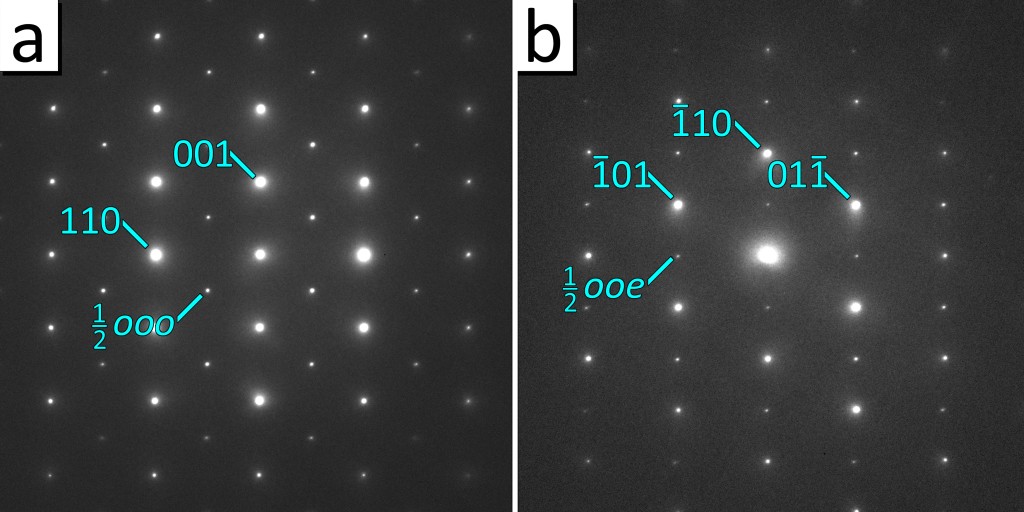}
\end{figure}

~
Here we present electron diffraction patterns from Ba$_{0.78}$Ca$_{0.22}$TiO$_3$ at room temperature as an example of a perovskite with oxygen octahedral tilting.  Fig.~\ref{SCT1} shows selected area electron diffraction patterns (SADPs) taken at the pseudocubic $\langle1\bar{1}0\rangle$ and  $\langle111\rangle$ zone axes.  At the $\langle1\bar{1}0\rangle$ zone axis, Fig.~\ref{SCT1}a), $\frac{1}{2}ooo$ spots indicate the presence of an anti-phase tilt system.  As is clear from Table \ref{summary}, it is more difficult to determine anti-phase tilt systems since they all produce $\frac{1}{2}ooo$ reflexions and are primarily distinguished by differences in their pattern of systematic absences.  Even in a crystal of moderate thickness like that used for Fig.~\ref{SCT1}a) multiple scattering `fills in' these absences, meaning that the observed reflexions could correspond to any anti-phase tilt system.  The $\langle111\rangle$ pattern of Fig.~\ref{SCT1}b) shows a single set of $\frac{1}{2}ooe$ spots, indicating the presence of a single in-phase tilt.  (We choose here to describe this as a $c^+$ tilt, but it would be equally valid to have chosen to describe it as $a^+$ or $b^+$.)

\begin{figure}
    \caption{Selected area electron diffraction patterns (SADPs) of $\langle100\rangle$ Ba$_{0.78}$Ca$_{0.22}$TiO$_3$.  a) A bright field image shows the existence of three types of domain. The locations of the SA aperture used to obtain SADPs b) and c) and d) are indicated.  Scale bar 500 nm.}
    \label{SCT2}
    \centering \includegraphics[width=0.8\columnwidth]{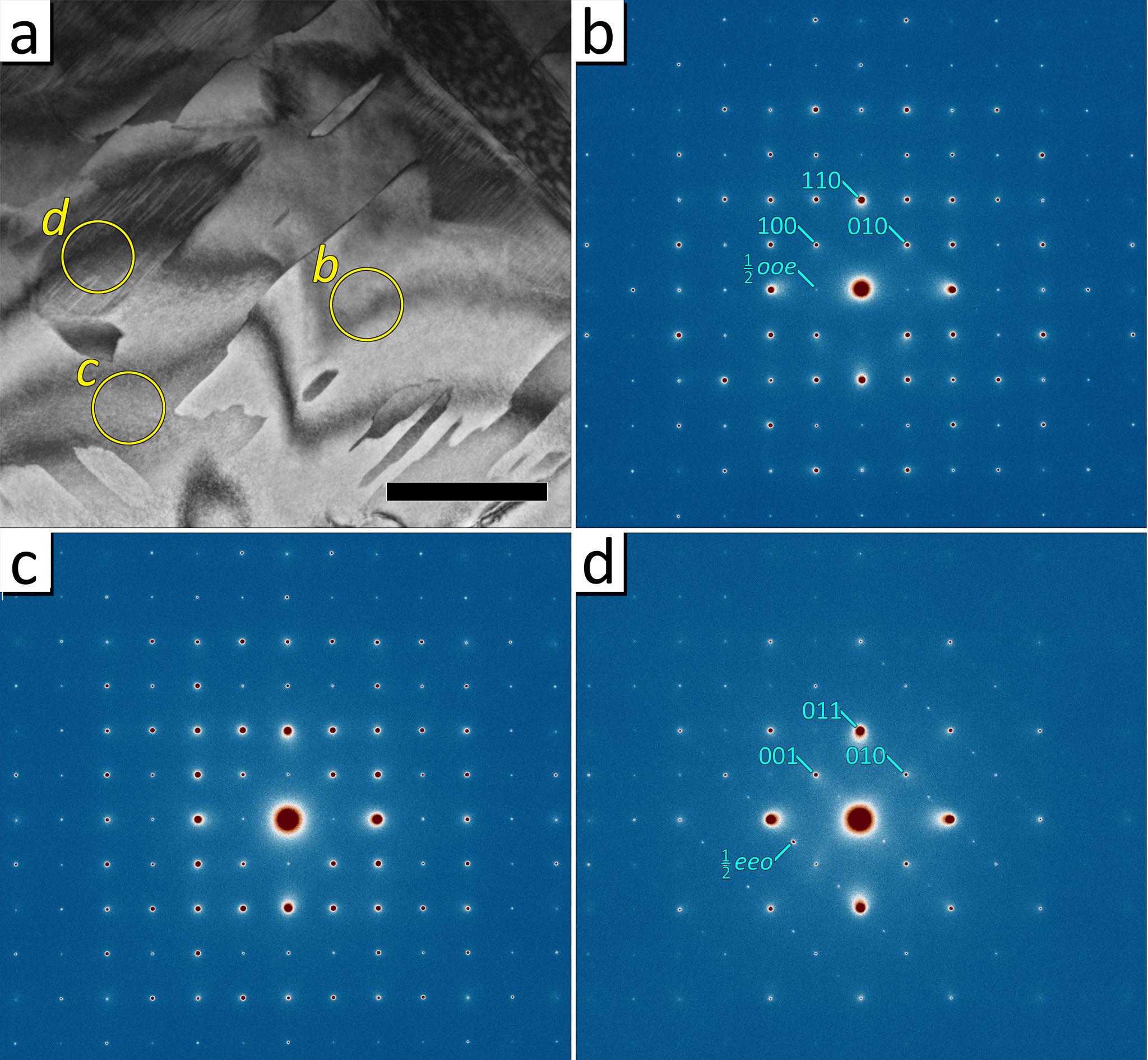}
\end{figure}

Fig.~\ref{SCT2}a) shows a bright field transmission electron microscope (TEM) image of a thin Ba$_{0.78}$Ca$_{0.22}$TiO$_3$ crystal viewed at the $\langle100\rangle$ axis. Three distinct domain types are visible, two of which have relatively uniform contrast and a third with a pattern of lines (these are anti-phase boundaries, with a preferred orientation of $(001)$).  Selected area electron diffraction patterns (SADPs) from the three domains are shown in Fig.~\ref{SCT2}b)-d).  Both Fig.~\ref{SCT2}b) and c) are $[001]$ patterns, based on our choice of $c^+$ in-phase tilt, and have $\frac{1}{2}ooe$ spots.  Reflexions along the horizontal and vertical rows passing through the direct beam are very much weaker -- suggesting they are systematically absent, but with a non-zero intensity resulting from multiple scattering.  This is consistent with the $|h| \neq |k|$ systematic absences in first-order superstructure reflexions produced by a $c^+$ tilt. Although the diffraction patterns b) and c) appear identical to the eye there is a very small deviation from a square geometry indicating an orthorhombic cell, as expected in a multi-tilt system with only one in-phase component.

  Fig.~\ref{SCT2}d) is taken from a domain with $[100]$ orientation.  Here, no half odd-odd-even spots are present, but the second-order `concert' $\frac{1}{2}eeo$ superstructure reflexions are readily visible, indicating a mixed tilt system.  Together, Figs.~\ref{SCT1} and ~\ref{SCT2} indicate that Ba$_{0.78}$Ca$_{0.22}$TiO$_3$ has a tilt system of either $Cmcm~a^0b^-c^+$, $Pnma~a^-a^-c^+$ or $P\frac{2_1}{m}~a^-b^-c^+$.

\section{Conclusions}

~

Equations governing the appearance of second order superstructure reflexions resulting from distortion modes in perovskites have been derived.  These reflexions are expected to be weaker than the first-order superstructure reflexions considered in \cite{B+S} as their kinematic intensity is proportional to $\delta^4$, where $\delta$ is the magnitude of the displacement of oxygen atoms from their position in the prototype perovskite structure.  The kinematic intensity of first-order superstructure reflexions is proportional to $\delta^2$. This may aid the interpretation of electron diffraction patterns and replicates the work of \cite{Glazer75} and \cite{Woodward}.  The emphasis on distortion mode, rather than space group, allows the interpretation of ED patterns without the need to rewrite vectors in real and reciprocal space for different unit cells.

\section{Acknowledgements}

~

We thank Alex Skaper and Sarujan Rupan for the data presented in Figs.~\ref{SCT1} and ~\ref{SCT2}.  This work was funded by EPSRC grant EP/V053701/1.

\newpage

\section{\textbf{Appendix}}

~

For completeness we compile in Table~\ref{summary} the conditions governing the existence of superstructure spots in the pseudo-cubic reference frame for the 14 crystallographically distinct Glazer tilt systems as listed by Howard and Stokes \cite{H+S}.  

~

Some general rules become apparent from Table~\ref{summary}.  The rules for in-phase tilting are quite straightforward, with each tilt system $a^+$, $b^+$, $c^+$ producing its own set of superstructure spots with pseudo-cubic indices $\frac{1}{2}eoo$, $\frac{1}{2}oeo$, $\frac{1}{2}ooe$ with no dependence on the presence of any other distortions.  Conversely, all antiphase tilt systems produce pseudo-cubic $\frac{1}{2}ooo$ superstructure spots, and are only distinguished by their systematic absences. Furthermore, because in these cases the type of superstructure reflexions is always the same, tilts of equal magnitude operating about different axes can result in changes to the set of systematic absences.  Accordingly, systematic absences are most apparent for the $a^-a^-a^-$ system.  This means that determining antiphase tilting systems is less straightforward than in-phase tilt systems.  For investigations using electron diffraction, it may thus be important to explore reciprocal space in three dimensions since systematic absences can readily be `filled in' by double diffraction where the possibility exists, particularly in the zero-order Laue zone.  Access to higher order Laue zones, or zone axes where no double diffraction pathways are present, is generally necessary.

Table~\ref{summary} shows that reflexions of the form $\frac{1}{2}eeo$, $\frac{1}{2}eoe$ and $\frac{1}{2}oee$ do not result from oxygen octahedral tilting.  They may, however, be produced by antiferrodistortive displacements of cations.  Calculation of extinction rules for these distortion modes is left as an exercise for the reader.\\
\\

\textbf{Equations - first order superstructure reflexions}

For each case in Table~\ref{summary} the polynomial equation governing the appearance of first order superstructure reflexions is given below.

1) $a^+a^+a^+$, $Im\bar{3}$
\begin{equation}
    \label{apapap}
    \begin{split}
    \frac{F_g}{2 \pi i f_g \delta} = & A \left(g_3 C - g_2 B \right)
    \left(1 + A^2\right)\left(1 - B^2\right)\left(1 - C^2 \right) \\
    & + B \left(g_1 A - g_3 C \right)
    \left(1 - A^2\right)\left(1 + B^2\right)\left(1 - C^2 \right) \\
    & + C \left(g_2 B - g_1 A \right)
    \left(1 - A^2\right)\left(1 - B^2\right)\left(1 + C^2 \right)
    \end{split}
\end{equation}
Conditions: $hkl=\frac{1}{2}ooe,~|h| \neq |k|; hkl=\frac{1}{2}oeo,~|h| \neq |l|;~hkl=\frac{1}{2}eoo,~|k| \neq |l|$\\

2) $a^0b^+b^+$, $I\frac{4}{m}mm$
\begin{equation}
    \label{a0bpbp}
    \begin{split}
    \frac{F_g}{2 \pi i f_g \delta} = B & \left(g_1 A - g_3 C \right)
    \left(1 - A^2\right)\left(1 + B^2\right)\left(1 - C^2 \right) \\
    + & C \left(g_2 B - g_1 A \right)
    \left(1 - A^2\right)\left(1 - B^2\right)\left(1 + C^2 \right)
    \end{split}
\end{equation}
Conditions: $hkl=\frac{1}{2}ooe,~|h| \neq |k|; hkl=\frac{1}{2}oeo,~|h| \neq |l|$\\

3) $a^0a^0c^+$, $P\frac{4}{m}bm$
\begin{equation}
    \label{a0a0cp}
    \frac{F_g}{2 \pi i f_g \delta} =  C \left(g_2 B - g_1 A \right)
    \left(1 - A^2\right)\left(1 - B^2\right)\left(1 + C^2 \right)
\end{equation}
Conditions: $hkl=\frac{1}{2}ooe,~|h| \neq |k|$\\

4) $a^0a^0c^-$, $\frac{4}{m}cm$
\begin{equation}
    \label{a0a0cm}
    \frac{F_g}{2 \pi i f_g \delta} =  C \left(g_2 B - g_1 A \right)
    \left(1 - A^2\right)\left(1 - B^2\right)\left(1 - C^2 \right)
\end{equation}
Conditions: $hkl=\frac{1}{2}ooo,~|h| \neq |k|$\\

5) $a^0b^-b^-$, $Imma$
\begin{equation}
    \label{a0bmbm}
    \begin{split}
    \frac{F_g}{2 \pi i f_g \delta} = & B \left(g_1 A - g_3 C \right)
    \left(1 - A^2\right)\left(1 - B^2\right)\left(1 - C^2 \right) \\
    + & C \left(g_2 B - g_1 A \right)
    \left(1 - A^2\right)\left(1 - B^2\right)\left(1 - C^2 \right)
    \end{split}
\end{equation}
Conditions: $hkl=\frac{1}{2}ooo,~k \ne l$ and $~!(|h|=|k|=|l|);~h~=~n, k~=~\pm~n~+~4m, l~=~\mp~n~+~4m, n, m, p$ integers.\\

6) $a^-a^-a^-$, $R\bar{3}c$
\begin{equation}
    \label{amamam}
    \begin{split}
    \frac{F_g}{2 \pi i f_g \delta} = & A \left(g_3 C - g_2 B \right)
    \left(1 - A^2\right)\left(1 - B^2\right)\left(1 - C^2 \right) \\
    + & B \left(g_1 A - g_3 C \right)
    \left(1 - A^2\right)\left(1 - B^2\right)\left(1 - C^2 \right) \\
    + & C \left(g_2 B - g_1 A \right)
    \left(1 - A^2\right)\left(1 - B^2\right)\left(1 - C^2 \right)
    \end{split}
\end{equation}
Conditions: $hkl=\frac{1}{2}ooo,~h \ne k~\&~k \ne l~\&~l \ne h, h=n, k=n+4m, l=n+4p,n, m, p$ integers.\\

7.  $a^+b^+c^+$, $Immm$
\begin{equation}
    \label{apbpcp}
    \begin{split}
    \frac{F_g}{2 \pi i f_g } = & A \delta_A \left(g_3 C - g_2 B \right)
    \left(1 + A^2\right)\left(1 - B^2\right)\left(1 - C^2 \right) \\
    + & B \delta_B \left(g_1 A - g_3 C \right)
    \left(1 - A^2\right)\left(1 + B^2\right)\left(1 - C^2 \right) \\
    + & C \delta_C \left(g_2 B - g_1 A \right)
    \left(1 - A^2\right)\left(1 - B^2\right)\left(1 + C^2 \right)
    \end{split}
\end{equation}
Conditions: $hkl=\frac{1}{2}ooe,~|h| \neq |k|; hkl=\frac{1}{2}oeo,~|h| \neq |l|; hkl=\frac{1}{2}eoo,~|k| \neq |l|$\\

8) $a^+a^+c^-$, $P\frac{4_2}{n}mc$
\begin{equation}
    \label{apbpcm}
    \begin{split}
    \frac{F_g}{2 \pi i f_g } = & A \delta_A \left(g_3 C - g_2 B \right)
    \left(1 + A^2\right)\left(1 - B^2\right)\left(1 - C^2 \right) \\
    + & B \delta_A \left(g_1 A - g_3 C \right)
    \left(1 - A^2\right)\left(1 + B^2\right)\left(1 - C^2 \right) \\
    + & C \delta_C \left(g_2 B - g_1 A \right)
    \left(1 - A^2\right)\left(1 - B^2\right)\left(1 - C^2 \right)
    \end{split}
\end{equation}
Conditions: $hkl=\frac{1}{2}ooo,~|h| \neq ~|k|; hkl=\frac{1}{2}oeo,~|h| \neq ~|l|; hkl=\frac{1}{2}eoo,~|k| \neq ~|l|$\\

9) $a^0b^+c^-$, $Cmcm$
\begin{equation}
    \label{a0bpcm}
    \begin{split}
    \frac{F_g}{2 \pi i f_g } = B & \delta_B \left(g_1 A - g_3 C \right)
    \left(1 - A^2\right)\left(1 + B^2\right)\left(1 - C^2 \right) \\
    + & C \delta_C \left(g_2 B - g_1 A \right)
    \left(1 - A^2\right)\left(1 - B^2\right)\left(1 - C^2 \right)
    \end{split}
\end{equation}
Conditions: $hkl=\frac{1}{2}ooo,~|h| \neq |k|;hkl=\frac{1}{2}oeo,~|h| \neq |l|$\\

10) $a^+b^-b^-$, $Pnma$
\begin{equation}
    \label{apbmbm}
    \begin{split}
    \frac{F_g}{2 \pi i f_g } = & A \delta_A \left(g_3 C - g_2 B \right)
    \left(1 + A^2\right)\left(1 - B^2\right)\left(1 - C^2 \right) \\
    + & B \delta_B \left(g_1 A - g_3 C \right)
    \left(1 - A^2\right)\left(1 - B^2\right)\left(1 - C^2 \right) \\
    + & C \delta_B \left(g_2 B - g_1 A \right)
    \left(1 - A^2\right)\left(1 - B^2\right)\left(1 - C^2 \right)
    \end{split}
\end{equation}
Conditions: $hkl=\frac{1}{2}eoo,~|k| \neq |l|; hkl=\frac{1}{2}ooo,~!(|h|=|k|=|l|),~k \ne l, (h=n, k=\pm n+4m, l=\mp n+4m), n, m$ integers.\\

11) $a^0b^-c^-$, $Cmcm$
\begin{equation}
    \label{a0bmcm}
    \begin{split}
    \frac{F_g}{2 \pi i f_g } = & B \delta_B \left(g_1 A - g_3 C \right)
    \left(1 - A^2\right)\left(1 - B^2\right)\left(1 - C^2 \right) \\
    + & C \delta_C \left(g_2 B - g_1 A \right)
    \left(1 - A^2\right)\left(1 - B^2\right)\left(1 - C^2 \right)
    \end{split}
\end{equation}
Conditions: $hkl=\frac{1}{2}ooo, ~!(|h|=|k|=|l|)$\\

12) $a^-b^-b^-$, $C\frac{2}{c}$
\begin{equation}
    \label{ambmbm}
    \begin{split}
    \frac{F_g}{2 \pi i f_g} = & A \delta_A \left(g_3 C - g_2 B \right)
    \left(1 - A^2\right)\left(1 - B^2\right)\left(1 - C^2 \right) \\
    + & B  \delta_B \left(g_1 A - g_3 C \right)
    \left(1 - A^2\right)\left(1 - B^2\right)\left(1 - C^2 \right) \\
    + & C  \delta_B \left(g_2 B - g_1 A \right)
    \left(1 - A^2\right)\left(1 - B^2\right)\left(1 - C^2 \right)
    \end{split}
\end{equation}
Conditions: $hkl=\frac{1}{2}ooo, ~!(|h|=|k|=|l|),~k \ne l$\\

13) $a^+b^-c^-$, $P\frac{2_1}{m}$
\begin{equation}
    \label{apbmcm}
    \begin{split}
    \frac{F_g}{2 \pi i f_g} = & A \delta_A \left(g_3 C - g_2 B \right)
    \left(1 + A^2\right)\left(1 - B^2\right)\left(1 - C^2 \right) \\
    + & B  \delta_B \left(g_1 A - g_3 C \right)
    \left(1 - A^2\right)\left(1 - B^2\right)\left(1 - C^2 \right) \\
    + & C  \delta_C \left(g_2 B - g_1 A \right)
    \left(1 - A^2\right)\left(1 - B^2\right)\left(1 - C^2 \right)
    \end{split}
\end{equation}
Conditions: $hkl=\frac{1}{2}ooo, ~!(|h|=|k|=|l|);~\frac{1}{2}eoo,~|k| \neq |l|$\\

14) $a^-b^-c^-$, $P\bar{1}$
\begin{equation}
    \label{ambmcm}
    \begin{split}
    \frac{F_g}{2 \pi i f_g} = & A \delta_A \left(g_3 C - g_2 B \right)
    \left(1 - A^2\right)\left(1 - B^2\right)\left(1 - C^2 \right) \\
    + & B  \delta_B \left(g_1 A - g_3 C \right)
    \left(1 - A^2\right)\left(1 - B^2\right)\left(1 - C^2 \right) \\
    + & C  \delta_C \left(g_2 B - g_1 A \right)
    \left(1 - A^2\right)\left(1 - B^2\right)\left(1 - C^2 \right)
    \end{split}
\end{equation}
Conditions: $hkl=\frac{1}{2}ooo, ~!(|h|=|k|=|l|)$\\
\\

\textbf{Equations - second order superstructure reflexions}

For each case in Table~\ref{summary} the polynomial equation governing the appearance of second order superstructure reflexions is given below.  In all cases the kinematical intensity of these reflexions is proportional to $\delta^4$ and they are expected to be weak in comparison with first-order superstructure reflexions, whose intensity is proportional to $\delta^2$.

1) $a^+a^+a^+$, $Im\bar{3}$
\begin{equation}
    \label{apapap2}
    \begin{split}
    \frac{F_g}{2 \pi^2 f_g} = & -g_1 g_2 AB 
    \left(1 - A^2\right)\left(1 - B^2\right)\left(1 + C^2 \right) \delta^2 \\
    & - g_2 g_3 BC
    \left(1 + A^2\right)\left(1 - B^2\right)\left(1 - C^2 \right) \delta^2 \\
    & - g_3 g_1 CA
    \left(1 - A^2\right)\left(1 + B^2\right)\left(1 - C^2 \right) \delta^2
    \end{split}
\end{equation}
Conditions: $hkl=\frac{1}{2}ooe,~h,k \neq 0; hkl=\frac{1}{2}oeo,~k,l \neq 0; hkl=\frac{1}{2}eoo,~l,h \neq 0$.  These sets of second order superstructure reflexions are the same systems as found for first order superstructure reflexions, but have fewer systematic absences.  No new systems of reflexions are present.\\

2) $a^0b^+b^+$, $I\frac{4}{m}mm$
\begin{equation}
    \label{a0bpbp2}
    \frac{F_g}{2 \pi^2 f_g} = - g_2 g_3 BC
    \left(1 + A^2\right)\left(1 - B^2\right)\left(1 - C^2 \right) \delta^2
\end{equation}
Conditions: $hkl=\frac{1}{2}eoo,~l,h \neq 0$.  This is an additional set of reflexions and completes the odd-odd-even sets, since first order systematic reflexions have the form $\frac{1}{2}ooe,~|h| \neq |k|;~\frac{1}{2}oeo,~|h| \neq |l|$.\\

3) $a^0a^0c^+$, $P\frac{4}{m}bm$ and 4) $a^0a^0c^-$, $\frac{4}{m}cm$

No second order superstructure reflexions are generated in tilted perovskites with a single tilt system.\\

5) $a^0b^-b^-$, $Imma$; 6) $a^-a^-a^-$, $R\bar{3}c$; 11) $a^0b^-c^-$, $Cmcm$; 12) $a^-b^-b^-$, $C\frac{2}{c}$; and 14) $a^-b^-c^-$, $P\bar{1}$

Structures with all anti-phase tilts do not generate visible second order superstructure reflexions, since they coincide with matrix reflexions.\\

7) $a^+b^+c^+$, $Immm$
\begin{equation}
    \label{apbpcp2}
    \begin{split}
    \frac{F_g}{2 \pi^2 f_g} = & -g_1 g_2 AB 
    \left(1 - A^2\right)\left(1 - B^2\right)\left(1 + C^2 \right) \delta_A \delta_B \\
    & - g_2 g_3 BC
    \left(1 + A^2\right)\left(1 - B^2\right)\left(1 - C^2 \right) \delta_B \delta_C \\
    & - g_3 g_1 CA
    \left(1 - A^2\right)\left(1 + B^2\right)\left(1 - C^2 \right) \delta_C \delta_A
    \end{split}
\end{equation}
Conditions: $hkl=\frac{1}{2}ooe,~h,k \neq 0; hkl=\frac{1}{2}oeo,~k,l \neq 0; hkl=\frac{1}{2}eoo,~l,h \neq 0$. These are essentially the same as for 1), $a^+a^+a^+$, but with differing kinematic intensities depending upon the relative magnitudes of the different tilts.\\

8) $a^+a^+c^-$, $P\frac{4_2}{n}mc$
\begin{equation}
    \label{apapcm2}
    \begin{split}
    \frac{F_g}{2 \pi^2 f_g} = & -g_1 g_2 AB 
    \left(1 - A^2\right)\left(1 - B^2\right)\left(1 + C^2 \right) \delta_A^2 \\
    & - g_2 g_3 BC
    \left(1 + A^2\right)\left(1 - B^2\right)\left(1 + C^2 \right) \delta_A \delta_C \\
    & - g_3 g_1 CA
    \left(1 - A^2\right)\left(1 + B^2\right)\left(1 + C^2 \right) \delta_C \delta_A
    \end{split}
\end{equation}
Conditions: $hkl=\frac{1}{2}ooe,~h,k \neq 0; hkl=\frac{1}{2}eoe,~h,l \neq 0; hkl=\frac{1}{2}oee,~h,l \neq 0$.  Like 2) $a^0b^+b^+$, the first-order superstructure reflections of type $\frac{1}{2}oeo$ and $\frac{1}{2}eoo$ are completed by the first set of odd-odd-even second-order reflexions.  The second and third sets of second-order reflexions, of even-even-odd type, complement them. \\

9) $a^0b^+c^-$, $Cmcm$
\begin{equation}
    \label{a0bpcm2}
    \frac{F_g}{2 \pi^2 f_g} = - g_2 g_3 BC
    \left(1 + A^2\right)\left(1 - B^2\right)\left(1 + C^2 \right) \delta_B \delta_C
\end{equation}
Conditions: $hkl=\frac{1}{2}eoe, k,l \neq 0$.  This set is complementary to the first-order superstructure reflexions of type  $hkl=\frac{1}{2}oeo$.\\

10) $a^+b^-b^-$, $Pnma$
\begin{equation}
    \label{apbmbm2}
    \begin{split}
    \frac{F_g}{2 \pi^2 f_g} = & -g_1 g_2 AB 
    \left(1 - A^2\right)\left(1 + B^2\right)\left(1 + C^2 \right) \delta_A \delta_B \\
    & - g_2 g_3 BC
    \left(1 + A^2\right)\left(1 + B^2\right)\left(1 + C^2 \right) \delta^2_B \\
    & - g_3 g_1 CA
    \left(1 - A^2\right)\left(1 + B^2\right)\left(1 + C^2 \right) \delta_B \delta_A
    \end{split}
\end{equation}
Conditions: $\frac{1}{2}oee,~h,k,l \neq 0$.  Note the second line gives no second-order reflexions.\\

13) $a^+b^-c^-$, $P\frac{2_1}{m}$
\begin{equation}
    \label{apbmcm2}
    \begin{split}
    \frac{F_g}{2 \pi^2 f_g} = & -g_1 g_2 AB 
    \left(1 - A^2\right)\left(1 + B^2\right)\left(1 + C^2 \right) \delta_A \delta_B \\
    & - g_2 g_3 BC
    \left(1 + A^2\right)\left(1 + B^2\right)\left(1 + C^2 \right) \delta_B \delta_C \\
    & - g_3 g_1 CA
    \left(1 - A^2\right)\left(1 + B^2\right)\left(1 + C^2 \right) \delta_B \delta_A
    \end{split}
\end{equation}
Conditions: $\frac{1}{2}oee,~h,k,l \neq 0$.  Similar to (10).\\

\newpage

\referencelist[references] 

\end{document}